\documentclass[conference]{IEEEtran}
\IEEEoverridecommandlockouts
\usepackage{graphicx}
\usepackage[english]{babel}
\usepackage{blindtext}
\usepackage{wrapfig}
\usepackage{listings}
\usepackage{balance}
\usepackage{booktabs}
\usepackage{xcolor}
\usepackage{amsmath}
\usepackage[tight]{subfigure}
\def\BibTeX{{\rm B\kern-.05em{\sc i\kern-.025em b}\kern-.08em
    T\kern-.1667em\lower.7ex\hbox{E}\kern-.125emX}}

\usepackage{float}

\usepackage{hyperref}

\newcommand\systemname{\textsc{ReactNet}\xspace}

\begin{document}

\title{A Demand-aware Networked System Using Telemetry and ML  with \systemname}

\author{\IEEEauthorblockN{Seyed Milad Miri}
\IEEEauthorblockA{\textit{TU Berlin} \\
Berlin, Germany}
\and
\IEEEauthorblockN{Stefan Schmid}
\IEEEauthorblockA{\textit{TU Berlin and Fraunhofer SIT}\\
Berlin, Germany}
\and
\IEEEauthorblockN{Habib Mostafaei}
\IEEEauthorblockA{\textit{Eindhoven University of Technology}\\
Eindhoven, Netherlands}
}






\maketitle

\begin{abstract}
Emerging network applications ranging from video streaming to virtual/augmented reality need to provide stringent quality-of-service (QoS) guarantees in complex and dynamic environments with shared resources. A promising approach to meeting these requirements is to automate complex network operations and create self-adjusting networks. These networks should automatically gather contextual information, analyze how to efficiently ensure QoS requirements, and adapt accordingly. This paper presents \systemname, a self-adjusting networked system designed to achieve this vision by leveraging emerging network programmability and machine learning techniques. Programmability empowers \systemname by providing fine-grained telemetry information, while machine learning-based classification techniques enable the system to learn and adjust the network to changing conditions. Our preliminary implementation of \systemname in P4 and Python demonstrates its effectiveness in video streaming applications.
\end{abstract}


\section{Introduction}\label{sec:intro}

Communication networks have become a critical infrastructure of our digital society, imposing increasingly stringent requirements on their dependability and performance. These requirements, however, stand in stark contrast to today's manual and error-prone approach to managing and operating networks, as well as the increasing complexity and scale of networks. Indeed, many communication networks today need to serve a wide spectrum of applications with different performance requirements. These applications typically share network resources in complex ways and have demand patterns that may be hard to predict.
For example, emerging applications such as online gaming, video streaming, or virtual/augmented reality may be latency-critical, while a distributed AI application may be bandwidth-hungry. In addition to the inherent complexity of meeting diverse requirements of the network applications, the efficient operation of such networks is further challenged by the limited visibility operators typically have into the current network traffic, as highlighted in prior research~\cite{reactNet-conext21,ray-osdi18}. The network operators need to develop several scripts to tailor the network to specific workloads, which are prone to bugs~\cite{self-drivingNetwrk-Feamster17}.

Automating the management and operation of communication networks is key to overcoming the complexities and dependability challenges of manual network operations~\cite{programmability-CCR20}. A particularly appealing vision is a fully self-adjusting network: a network that automatically measures itself, gathers information about its context and environment, current demands, and loads, to then evaluate the most efficient and effective resource allocation to meet quality of service (QoS) requirements. Specifically, we are interested in self-adjusting networks that continuously measure, analyze, and adapt.

According to~\cite{kalmbach2018empowering}, the algorithms in self-adjusting networks must continuously be updated as the requirements and demands change frequently, and usually, they vary from one system to the other. The study also notes another challenge: the cost and risk calculation for each system if it encounters an unknown problem as humans do not manage it. So every aspect of the possible error should be considered, depending on which method we are using the self-driving 

Programmable networks can play a vital role in the realization of self-adjusting systems, as they provide flexibility to collect fine-grained telemetry information about the network traffic flows and adapt the forwarding rules accordingly at line rate without delaying the packets of the flows~\cite{kellerer2019adaptable}. To achieve the adaptation goal self-adjusting networks, steering traffic, and controlling connections among the endpoints are not enough since we should also consider the storage and processing capabilities of each compute element, particularly those responsible for adaptation and adjustment~\cite{laghrissi2018survey, hantouti2018traffic}. Specifically, self-adjusting networks should dynamically observe their current state and automatically react to optimize for specific performance goals accordingly~\cite{reactNet-conext21}.
For example, some Internet customers cannot bear a connection with jitter and latency, which could negatively impact their systems or services. Furthermore, operators can leverage machine learning (ML) algorithms to automate parts of network management and simplify administration. By inspecting packets and flows through data plane programmability~\cite{p4-ccr14}, they can perform further analysis and predict network behavior in various situations using ML methods. ML enables networked systems to adapt to different conditions and respond automatically based on trained data. Additionally, operators use ML algorithms to classify network traffic, balancing resource consumption in hardware devices with achieving reasonable classification accuracy~\cite{kellerer2019adaptable}.

This paper presents \systemname~\footnote{The preliminary version of this work has been published in~\cite{reactNet-conext21}.}, a self-adjusting networked system that aims to realize the vision of self-adjusting networks. \systemname is enabled by the increased flexibility of communication networks today, particularly network programmability: e.g., programmable switches empower \systemname through fine-grained telemetry information.
\systemname relies on machine learning-based classification techniques, allowing the system to learn and adjust the network to the new conditions. 
Hence, \systemname can learn from the ongoing network conditions and adapt to the new state according to the desired QoS and Quality of Experience (QoE) needs.
Our system measures the packet processing time of the desired flows and can set a threshold for them when the packet enters the network. This feature gives the system a powerful mechanism to tune the network application needs to the desired QoS or QoE requirements after applying the ML logic. In contrast, conventional capacity over-provisioning techniques lack such dynamicity in meeting the application needs.

We report on a prototype implementation of \systemname in a programmable data plane, i.e., P4~\cite{p4-ccr14}, and Python,
and also present a preliminary performance evaluation with case studies.
Our prototype evaluation on a video streaming scenario shows that by adapting, \systemname can indeed meet the QoE requirements of the application. We also test the accuracy of our ML classifiers on a trace of packets of IoT devices~\cite{sivanathan2018classifying} and observe an accuracy of 99\% for the classified packets. 

The remainder of this paper is organized as follows. We provide a preliminary discussion about the role of traffic classification in Section~\ref{sec:traffic}.
Section~\ref{sec:architecture} presents the design of \systemname. The proof-of-the-concept comes in Section~\ref{sec:proof}. Simulation results are illustrated in Section~\ref{sec:results}. In Section~\ref{sec:soa}, we survey the related literature. 
Finally, Section~\ref{sec:concl} concludes the paper.

\section{Network Traffic Classification}
\label{sec:traffic}

Network traffic classification has become a crucial part of any networked system. Network operators seek solutions to classify network traffic to address various issues~\cite{shafiq2016network}. The network administrators need to monitor flows within their networks to take appropriate actions according to the underlying network traffic, ensuring meeting the application requirements~\cite{nguyen2008survey}. Today, by analyzing flows and packets, we can identify distinct patterns, find correlations between features, and pinpoint failures within the network. Additionally, network traffic analysis offers benefits such as intrusion detection and achieving optimal QoS~\cite{pacheco2018towards}.

Initially, the concept of network traffic flows has been crucial for intrusion detection. In \cite{frank1994artificial}, a collection of data was used to analyze and classify ML methods. The algorithms used in ML should address problems that evolve and change over time, requiring continuous updates by technical experts~\cite{frank1994artificial}. 
The initial application of ML algorithms for traffic classification and intrusion detection in networks was in 1994. Generally, ML methods utilize features from a dataset as inputs to identify and illustrate patterns between features with different characteristics. After learning and recognizing these patterns, the output describes these patterns and structures~\cite{nguyen2008survey}.

Features are distinguishable elements of network traffic that can be identified in unknown IP traffic. These features include properties of IP packets or flows, such as protocol, flow duration, IP addresses, and source or destination ports. We use these features in ML algorithms for training on known datasets to analyze and classify data, detecting feature correlations among flows or packets. The algorithm then uses the trained data to classify other unknown data.

Supervised and unsupervised learning-based methods are the most applied ML techniques for traffic classification in the networked systems~\cite{MLSurvey-IEEESurvey18}. In supervised learning, the algorithm processes data based on a predefined set of labels and requires preprocessing of the dataset~\cite{challengesofLabeling-NetAI21}. Conversely, in unsupervised learning, ML algorithms analyze the dataset and find patterns without prior modification or preprocessing, generally grouping features into clusters~\cite{berry2019supervised}. However, to classify the packets of different flows belonging to various applications, we use supervised learning-based methods in this paper~\cite{ML-CCS22}.

\section{\systemname Architecture}\label{sec:architecture}

\systemname has three main components: 1) Collecting valuable data from ongoing traffic, 2) learning the network status from collected data, and 3) adjusting the network based on the learned situation. We now explain these components and how \systemname reaches the design goals.

\subsection{Data Collection}
Collecting data from the ongoing network traffic is a key operation of self-adjusting networks. The legacy approach to get insights from the network traffic is to use sampling. This technique efficiently collects sample packets from the ongoing traffic flow and provides partial information. Nevertheless, the sampling technique adds non-negligible costs to the network since it requires extra computing power to analyze the data. However, leveraging the programmable hardware~\cite{p4-ccr14} can greatly support efficient data collection. Programmable switches can provide insights about packets of all flows without degrading the performance of the network. Therefore, we build \systemname on programmable networks for data collection.

\begin{figure*}[tp]
    \centering
    \includegraphics[width=0.75\linewidth]{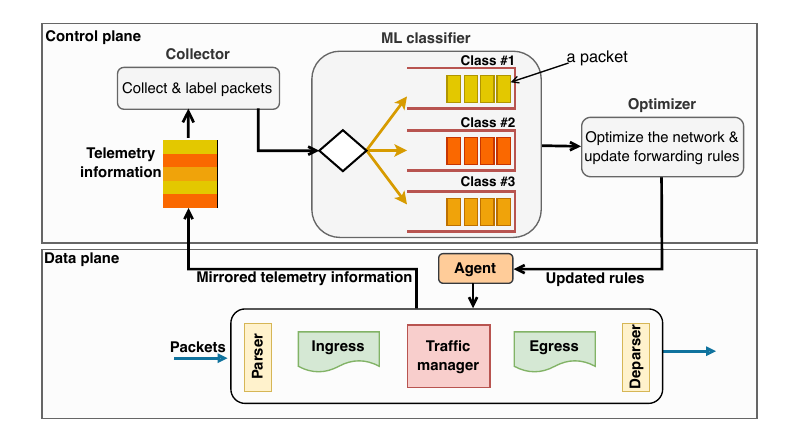}
    \caption{The architecture of \systemname for self-adjustable networks, implemented on a P4 programmable switch. The data for training, i.e., packet headers and fine-grained telemetry information, of the system is collected via the data plane, while the classification of packets ( to three classes in this figure), optimizing, and updating the network are performed via the control plane. }
    \label{fig:architecture}
\end{figure*}

We can collect insights from the packets by looking at the packet header, such as source and destination IP addresses, source and destination ports, and protocol number. However, the insights collected from the packets of different flows can be significantly improved using the In-band Network Telemetry (INT) provided by P4~\cite{int-spec}. Examples of such telemetry information are queue occupancy of network devices along the path from source to destination, packet inter-arrival time, and packet processing time. 

The network operators of \systemname can adjust data collection depending on the time of need using a customized flag. We use the mirroring feature of the programmable devices to collect data from the traffic flows. If the flag is set, the \systemname mirrors the traffic to the designated egress port towards the collector. Otherwise, the traffic flows follow the forwarding rules for the routing decisions. The designed system can also mirror the traffic according to the desired interval using a timer if the information of the packets is unnecessary. This feature can help the system save bandwidth and decrease the overhead of handling all the packets. We use P4 registers to store the flag value and the timer. 

\systemname can also collect datasets from the network for different purposes since our data collection mechanism for adjusting the network is generic. We use \textit{influxDB} to store the mirrored telemetry data in a database.

\subsection{Learning from Data}
The second pillar of a self-adjusting network is the capability to learn from the ongoing network conditions. It empowers the system to make decisions without human interaction based on experience. 

One common way to learn from data is to use ML learning techniques. For example, we can use these techniques to classify packets of different flows collected previously. Some widely used classifiers in networking~\cite{ML-syscon18} are Decision Trees, Support-Vector Machine (SVM), Random Forest (RF), and K-Nearest Neighbor (KNN). These classifiers are supervised learning techniques that use labeled datasets to train the system and predict outcomes accurately. The classifiers of \systemname need the label information from the network operators to classify the incoming packets accordingly. \systemname labels the data before mirroring them. Our system can learn from the trained dataset and properly decide on upcoming conditions by checking the precise information collected from the packets.

\subsection{Adjusting the Network}
The self-adjustable network adapts itself to a new state after learning from the current state of the system. This adaption can be accomplished in several ways, such as updating the forwarding rules to balance the traffic of different links or adjusting the priority of the flows to state a few. For instance, consider a scenario where a network should adapt itself to forward the packets of business transactions without delaying them when competing for network resources. The system can learn from the collected data and the packet processing time of the packets if it needs to take a reaction to the ongoing flows. The network operators can specify the requirements of the applications to adjust the network. 

The current implementation of \systemname adjusts the network flows by assigning the desired priority for the packets of different flows. However, this is a design choice rather than a system limitation.  \systemname can also forward the traffic toward multiple links to balance the load of the network. This feature of our system needs to be tuned according to the application's needs. For example, we test the system for a video-streaming scenario to check the impact of self-adjusting on the video quality when the network has to carry non-responsive UDP traffic in \textsection\ref{sec:results}.

\systemname adjusts the network-- via the proper API of the programmable switches-- by updating the priority of the flows to handle the upcoming traffic based on the detected traffic pattern. The forthcoming traffic of the tuned flow follows the new pattern to improve the QoS or QoE. The network operator of \systemname can also set the flag of mirroring the packets via the switch API whenever the dataset needs updating with the new telemetry information.

\subsection{Bringing It All Together}
We now explain the architecture of our system by putting all the mentioned properties together. \systemname comprises two parts implemented in programmable networks' data and control planes. The data plane includes at least one P4 programmable switch, which can be the switch of the access network, and provides all the means to our architecture. In contrast, the system control plane gets the applications' needs as the input and adapts the forwarding rules accordingly. We name the components of \systemname as follows. \textit{P4 switch}, \textit{collector}, \textit{ML classifier}, and \textit{Adjuster}.

Fig.~\ref{fig:architecture} shows the architecture of \systemname. The \textit{P4 switch} forwards the incoming packets to the designated egress ports according to the forwarding rules. If the corresponding flag to the mirror is set, the switch also appends the telemetry information into the packets. Then, it mirrors the packet with the desired header fields, including telemetry information to the collector. The \textit{collector} receives the mirrored packets and puts them into the database by adding the proper label to each packet. We need data labeling to train our system using supervised machine-learning techniques. \systemname gets the label information from the control plane and stores it in a proper data structure in the data plane. The \textit{ML classifier} of \systemname reads the data from the database and classifies the packets into different classes according to the network policy. Then, it sends the classified packets to \textit{Adjuster} component that optimizes the rules and updates them accordingly. The P4 runtime agent updates the rules on the switch, and upcoming traffic flows will follow the updated rule.

Our system measures the packet processing time of the desired flows and can set a threshold for them when the packet enters the network. This feature empowers the system with a more sophisticated mechanism to tune the network application needs to the desired QoS or QoE requirements. While the conventional capacity over-provisioning techniques lack such dynamicity in meeting the application needs.

\section{Proof-of-Concept}\label{sec:proof}
We implement \systemname in P4 using the Behavioral Model v2 (BMv2) switch in Mininet and the ML part in Python.

\noindent\textbf{Flow identification.}
\systemname needs to identify the packets of different flows after classification by the ML techniques. We assign an ID for each flow and use P4 registers to track them. We also define a register called \texttt{prio\_reg} to set the priority of the packets for different flows. Since the amount of available memory on the programmable device is limited, the network operators of \systemname can prioritize the IDs based on the application need and service level agreements after adjusting the network. The switch forwards the packets based on their ID and corresponding priority value. 

\begin{table}[tp]
\centering
\caption{Feature used to build our dataset for the ML-based classification.}
\begin{tabular}{lc}
\toprule
\textbf{Feature} & \textbf{Size in bit} \\
\midrule
 Ingress\_port & 9 \\
 Flow interval time &  48 \\
 enq-qdepth & 19 \\
 deq-qdepth & 19 \\
 deq-timedelta & 32 \\
 Protocol number & 8 \\
 Source port & 16 \\
 Destination port & 16 \\
 IPv4 source address & 32 \\
 IPv4 destination address & 32 \\
\bottomrule
\end{tabular}
\label{tab:dataset}
\end{table}

\noindent\textbf{Data collection.}
We implement the data collection part of our system from the network flows in P4 in the egress control flow since we have access to all telemetry information. Table~\ref{tab:dataset} shows the telemetry information with their size in bits \systemname extracts from every packet. We explain some of the features that need more clarification. \textit{Flow interval time} specifies how long each packet spends between the ingress and egress ports. \textit{enq-qdepth} shows the depth of the queue when the packet was enqueued. \textit{deq-qdepth} specifies the depth of the queue when the packet was dequeued. \textit{deq-timedelta} is the time that the packet was in the queue. We append the telemetry information to the packet and forward it to our data collector.

\noindent\textbf{ML classification.}
We implement the ML classification component of \systemname using the scikit-learn library in Python. The library has a reach set of implementations for different ML techniques. Our system applies different supervised learning techniques to classify the packets. The user can also specify the desired classification algorithm to apply to the collected data. We use the recommendation of~\cite{ml-classification-HotNet19} to provide the required label information for packet classification.

\noindent\textbf{Self-tuning dataset.}
\systemname by default collects telemetry information of each packet. However, the network operator can tune our system to collect the amount of the cloned packets using a predefined time interval. The time interval information is an input for the system provided by the control plane. This feature avoids the overloading of the collector by capturing many packets. 

We define a dedicated timer for each flow using P4 registers to measure the packet processing time in the switch and implement it as follows. We store the current timestamp of the packet in the register and then subtract this value from the timestamp of the next packet. By comparing the subtracted result with an arbitrary threshold, the switch decides to either clone or forward the packet of that flow.

\systemname exploits the cloning mechanism of programmable switches to send a copy of the telemetry information of each packet. The switch forwards the cloned packets via the egress port connected to the switch. However, having a direct link connection from the switch to the collector is unnecessary since we can modify the packet header to reach the collector according to its destination IP address. 

We use Logstash~\cite{Logstash} to filter the received packets by providing the key elements or features in the configuration file. \systemname stores the filtered data sent from Logstash into the Influx database. We use the Influxdb plugin of Logstash to send the data to Influxdb.

\noindent\textbf{Adjusting the network by changing the priority of the packets.} We use a set of registers to set the priority for different packets. \systemname updates the values of these registers via \textit{simple\_switch} API. We check the value of the corresponding register to set the proper priority for the packet.

\section{Performance Evaluation}\label{sec:results}
This section reports the performance of \systemname when applied to a video streaming scenario. We also test the performance of the ML classifiers on an IoT trace, including $\approx$ 600k packets.

\begin{figure}[tp]
  \begin{center}
    \includegraphics[width=\linewidth]{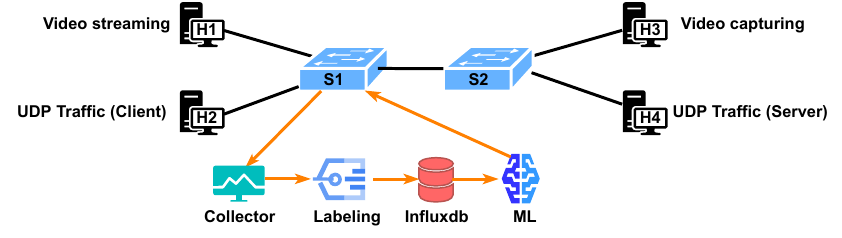}
  \end{center}
  \caption{Our evaluation network topology for the video streaming application.}
  \label{fig:topo}
\end{figure}

\noindent\textbf{Testbed network.}
We create a dumbbell topology with two P4 switches in Mininet connected with a 2Mbps link. We attach two hosts to switch \textit{S1} to generate the traffic, namely, \textit{H1} for video streaming and \textit{H2} for UDP traffic. We generate UDP traffic using \textit{Iperf}. The two hosts, i.e., \textit{H3} and \textit{H4}, connected to switch $S2$ receive traffic from the corresponding senders (see Fig.~\ref{fig:topo}). We also attach another host, i.e., \textit{collector}, to the switch \textit{S1} to receive the cloned packets.

\subsection{Video Streaming Scenario}
This experiment aims to show the capability of \systemname to adjust itself for a better QoE for the video streaming application.
We stream the ''Big Buck Bunny'' video three times using FFmpeg~\cite{ffmpeg}. We first stream the video without sending any background traffic. This experiment aims to obtain the QoE performance metrics as the ground-truth value for comparison. Then, we stream the video with background traffic without adjusting the network, i.e., without \systemname. Finally, we stream the video with background traffic and apply \systemname, i.e., with \systemname. The UDP client H2 sends 2Mbps traffic to the corresponding receiver host, i.e., H4, after 10 seconds in both scenarios.

We use Logstash to collect and label the cloned packets from the streaming video and UDP traffic. We also tune the Logstash configuration to a higher timer precision to get all cloned packets.
We apply higher priority to the packets of the video traffic in S1, while the priority of other flows remains at their default value.

\noindent\textbf{Impact on the total frame rate.} The frame rate of the original streaming video is 30 frames per second (FPS). The metric for the scenario without applying \systemname is 26.11 FPS, while with \systemname, it is 28.92 FPS.

\noindent\textbf{Impact on the image quality metric.}
We report the Peak Signal-to-Noise- Ratio (PSNR) as the main image quality metric. PSNR indicates the ratio between the maximum possible value of a signal and the power of distorting noise that affects the quality of that image~\cite{imageQuality-ICPR10}. A higher PSNR value for the quality of images in video streaming applications is preferred. 
For the reference image f and the disported image g, with the size of $M\times N$, the PNSR in~\cite{hore2010image} is defined as follows.

\begin{equation}
\text{PSNR}(f,g)=10\log_{10}(255^2/\text{MSE}(f,g)) 
\end{equation}

whereas:

\begin{equation}
\text{MSE}(f,g)= \frac{1}{MN}\sum_{i=1}^{M}\sum_{j=1}^{N}(f_{ij} - g_{ij})^2
\end{equation}

\begin{figure}[tp]
  \begin{center}
    \includegraphics[width=\linewidth]{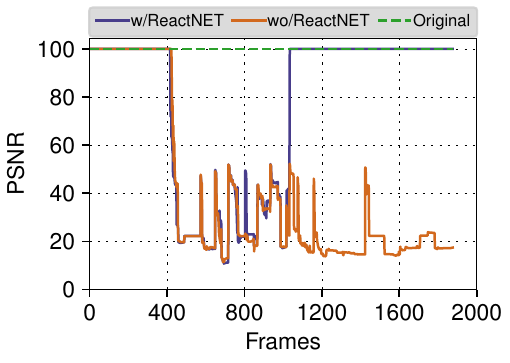}
  \end{center}
  \caption{The effect of \systemname on the PSNR metric of the streamed video.}
  \label{fig:psnr}
\end{figure}

Fig.~\ref{fig:psnr} shows that after adjusting the traffic rate of the video stream, the PSNR value of the streaming video significantly improves since the \systemname sets the higher priority to those packets. The main reason for such an improvement in PSNR values relies on adjusting the priority of the streaming packets on the switch by using \systemname.

\begin{figure}[tp]
  \begin{center}
    \includegraphics[width=\linewidth]{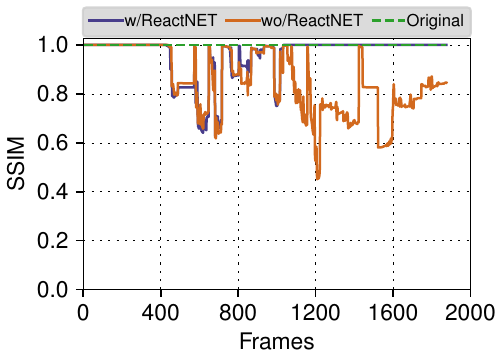}
  \end{center}
  \caption{The effect of \systemname on the SSIM metric of the streamed video.}
  \label{fig:ssim}
\end{figure}

Fig.~\ref{fig:ssim} shows the Structured Similarity Index Method (SSIM) metric of the streamed video.
The SSIM is a full-reference quality metric that checks the similarity of the reference and the test images by combining three factors: loss of correlation, luminance distortion, and contrast distortion. The value of SSIM varies in the range of [0,1], and zero shows no correlation or similarity, and the value of 1 indicates both images are identical. The streamed video achieves better SSIM when applying \systemname.

\begin{figure}[tp]
  \begin{center}
    \includegraphics[width=\linewidth]{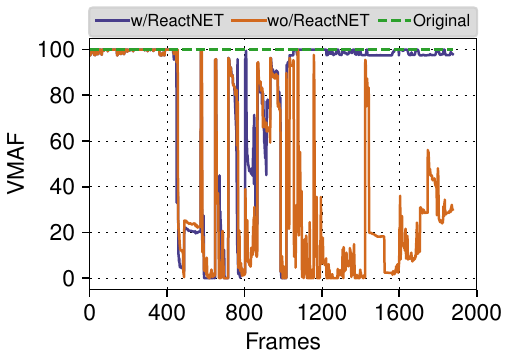}
  \end{center}
  \caption{The impact of using \systemname on the VMAF metric of the streamed video.}
  \label{fig:vmaf}
\end{figure}

\begin{figure*}[tp]
    \centering
    \subfigure[Video output with background traffic\label{fig:scene-before-ML}]{\includegraphics[width=.485\linewidth]{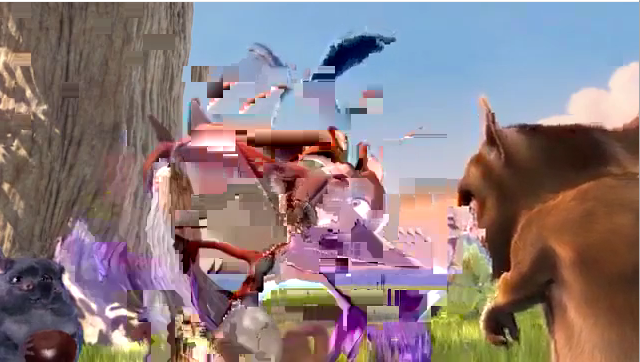}}
    \hfill
    \subfigure[Video output after using \systemname\label{fig:scene-after-ML}]{\includegraphics[width=.485\linewidth]{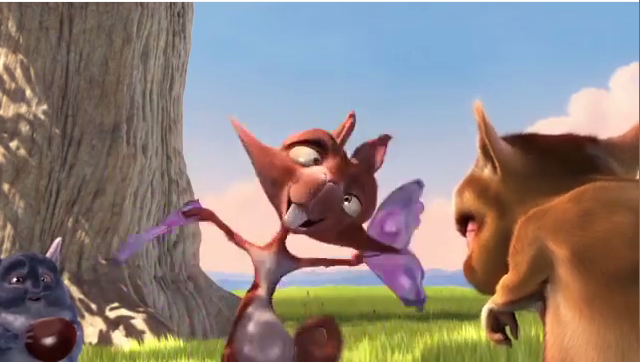}}
    \caption{Comparison of the identical scene of the two video outputs and the impact of \systemname on video quality.}
    \label{fig:FCT-WS-varyRS}
\end{figure*}

\noindent\textbf{Impact on the video quality metric.}
Netflix introduced Video Multi-Method Assessment Fusion (VMAF)~\cite{VMAF} as a video quality metric. VMAF assesses video quality after re-scaling and compression to detect degradation. The VMAF score ranges from 0 to 100, with higher values indicating better quality. As shown in Fig.\ref{fig:vmaf}, after adjusting the traffic priority for video streaming, the VMAF score of the stream closely approximates the ground-truth value.

Figures~\ref{fig:scene-before-ML} and \ref{fig:scene-after-ML} compare the identical frame of the "Big Buck Bunny" for the video outputs with background traffic for scenarios without and with \systemname. The receiver loses many frames before applying the ML classifier with \systemname resulting in suboptimal video quality for human observation.

\subsection{Accuracy of ML Classifiers of \systemname}
As the second application of our designed self-adjusting system, this section reports the accuracy of ML classifiers in classifying the packets of different flows. We use the traffic trace of IoT devices~\cite{sivanathan2018classifying} and use Tcpreplay to inject the packets into the network from host H2 in Fig.~\ref{fig:topo}. The trace contains $\approx$ 600k packets, and we replay them according to the capacity of the link between switch $S1$ and switch $S2$ in our topology.
Table~\ref{table:classes} shows the name of each class with the assigned number that we use in the classifiers. 

\begin{table}[h!]
\centering
\begin{tabular}{l c} 
 \hline
 Name & Class \\ [0.8ex] 
 \hline
 Energy & 0 \\
 Appliances & 1 \\
 Hubs & 2 \\
 Health-Monitors & 3 \\
 Cameras  & 4 \\
 Others  & 5 \\
 
\hline
\end{tabular}
\caption{The IoT class names and their assigned numbers in the classifiers}
\label{table:classes}
\end{table}

We use the port numbers of different applications in the trace to label the mirrored data in our dataset. The main reason for such a choice is that IoT devices use a few port numbers compared with non-IoT devices.
In addition, the devices made by the same manufacturer tend to use standard port numbers. This choice made our classification based on the IoT signaling pattern, which helped us easily classify the flows. We have five different classes, namely, energy, appliances, hubs, health-monitor, cameras, and others that indicate non-IoT devices.
We used three models of ML methods, namely k-Nearest Neighbour (KNN), Decision
Tree (DT) and Random Forest (RF) on the collected IoT data-trace to make a classification towards reaching optimal results.
Table~\ref{tbl:scores} shows that the decision tree and KNN-based classification have the most accurate results with 99\% accuracy.
\begin{table}[ht]
	\caption{The accuracy of different ML classification algorithms of \systemname on IoT trace.} \label{tbl:scores}
    \centering
 \resizebox{\linewidth}{!}{
	\begin{tabular}{lcccc}
		\toprule
	\textbf{Model} & \textbf{Accuracy} &\textbf{F1\_score} &\textbf{MSE}&\textbf{Precision}\\
		\midrule
		Decision Tree  & 0.99 &0.1 & 0.06 &0.75  \\
		K-Nearest Neighbors & 0.99 & 0.99& 0.001&0.87  \\
		Random Forest & 0.98 & 0.99& 0.11 &0.57 \\
		\bottomrule
	\end{tabular}
	}
\end{table}

\begin{figure*}[tp] 
    \centering
    \subfigure[DT ]{%
        \includegraphics[width=0.325\linewidth]{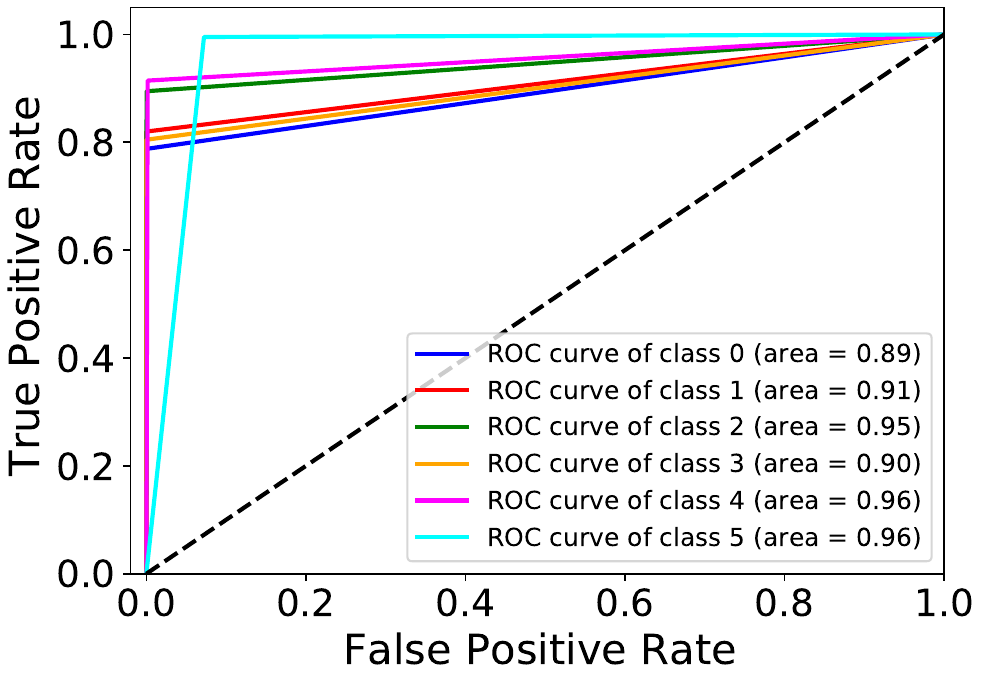}%
        \label{fig:DTroc}%
        }%
    \subfigure[KNN]{%
        \includegraphics[width=0.325\linewidth]{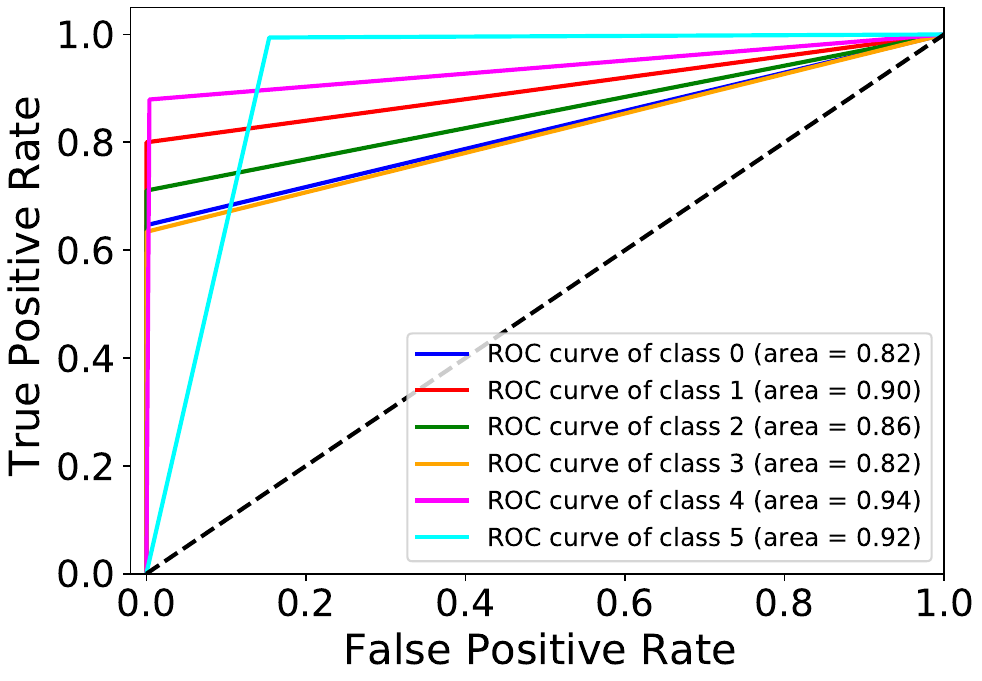}%
        \label{fig:KNNroc}%
        }%
    \subfigure[RF]{%
        \includegraphics[width=0.325\linewidth]{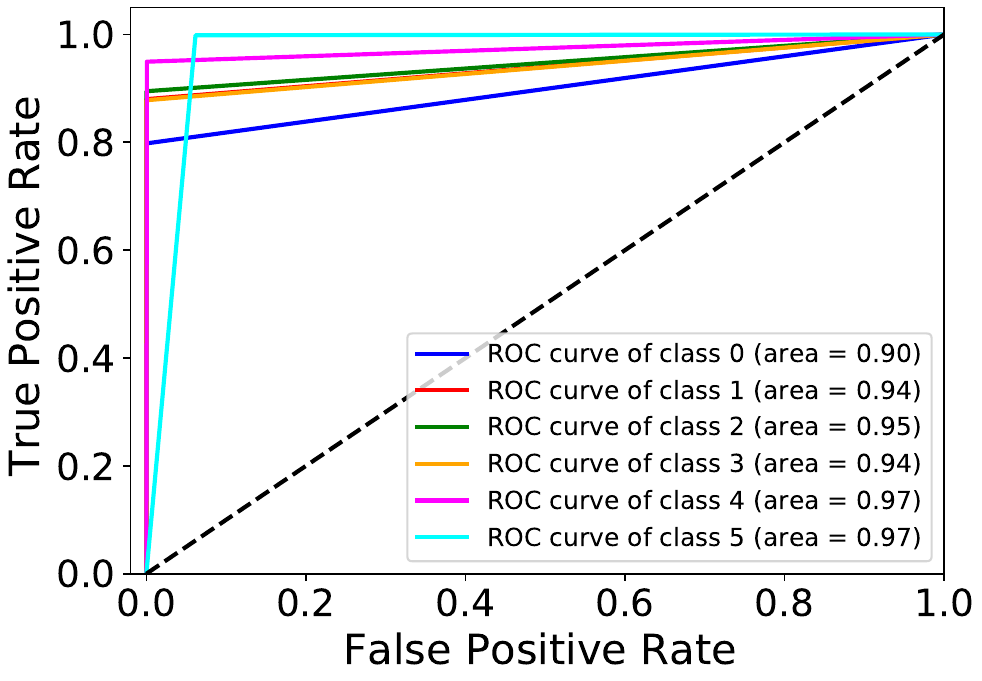}%
        \label{fig:RFroc}%
        }%
    \caption{ROC curve results for different classifiers for five classes in the dataset. }
    \label{fig:ROC}
\end{figure*}

We plot the Receiver Operating Characteristic (ROC) curve for DT, KNN, and RF packet classifiers in Figure~\ref{fig:ROC}. The ROC curve shows the True Positive Rate (TPR) against the False Positive Rate (FPR) for each class on the y and x-axis respectively for
different threshold~\cite{hoo2017roc}.
Fig.~\ref{fig:DTroc} shows the ROC curves for each class in~\ref{table:classes}. For each class in the DT methods, the ROC curves lie above the diagonal line x=y. Notably, the minimum AUC for class zero (Energy) is 0.89, indicating that the model can effectively distinguish between positive and negative classes. The~ROC for KNN methods for each class in Figure~\ref{fig:KNNroc} show that the AUC for classes 0 (Energy) and 3 (Health-Monitors) is 0.82. Lastly, the ROC  for RF in Figure \ref{fig:RFroc} demonstrates superior performance compared to the DT and KNN methods. The highest accuracy is observed for classes 4 (Cameras) and 5 (Others), with an AUC of 0.97.

\section{Related Work}\label{sec:soa}

The vision of self-adjusting and ``self-driving'' networks
has recently received much attention~\cite{self-drivingNetwrk-Feamster17,demand-aware-CCR19,kellerer2019adaptable}. It is enabled by the increasing programmability and flexibility of networks~\cite{programmability-CCR20} as well as the success of AI in various domains. Indeed, softwarization facilitates a more automated administration, operation, and management of the networked systems, and monitoring~\cite{zerwas2019netboa,kim2015band,safa-toit23}. 



We review some research and studies in the self-driven network area.
ReNet~\cite{avin2019renets} is a self-adjustable approach to optimizing route lengths in demand-aware networks (DANs). 
ReNet uses splay trees~\cite{sleator1985self}, a self-adjusting Binary Search Tree (BST), to adapt the network based on optimizing topology to facilitate routing issues. The system in~\cite{self-driving-Access21} proposes a self-driving management system based on intents to reduce the complexity of network management.

AI-based approaches such as NetBOA~\cite{zerwas2019netboa} also generally allow measuring and estimating critical system information such as CPU performance or network latency. A deep reinforcement learning-based approach to coordinate microservices in self-driving networks is proposed in~\cite{serviceCoordination-CNSM20} to manage them based on traffic patterns.  

The usage of ML has also been considered in the context of data planes~\cite{verification-NetAI19,switchML-nsdi21,reactNet-conext21}. For example, SwitchML offloads the distributed parallel training of part of machine learning systems to the network to reduce the amount of exchanged information and speed up their processing requirements using programmable networks~\cite{switchML-nsdi21}. \systemname currently relies on an external entity to run the ML classification task. We can offload the classification task of \systemname to the programmable switches. However, a careful architecture design needs to be considered due to the memory limitations of programmable switches.

\section{Conclusion}\label{sec:concl}
This paper introduced \systemname, a self-adjustable network that can adapt to the application requirements given by the network operators. Our system is built on two key enabler technologies: programmable networks and machine learning. Leveraging programmable networks enables the system to get telemetry information from all ongoing packets without delay. This provides more accurate data to our machine learning-based classification algorithms. Our evaluations showed that the system could tune the network to meet the QoE requirements for video streaming applications. Also, the machine learning techniques are highly accurate in classifying the packets of different applications. We plan to extend \systemname by adding more sophisticated strategies to optimize the network and leverage available resources. 

\balance

\section*{Acknowledgment}
This work was partially funded by the German Ministry for Education and Research as BIFOLD - Berlin Institute for the Foundations of Learning and Data (ref. 01IS18025A and ref. 01IS18037A),
as well as by German Research Foundation (DFG) project ReNO (SPP 2378), 2023-2027.

\bibliographystyle{IEEEtran}
\bibliography{paper}

\begin{thebibliography}{10}
\providecommand{\url}[1]{#1}
\csname url@samestyle\endcsname
\providecommand{\newblock}{\relax}
\providecommand{\bibinfo}[2]{#2}
\providecommand{\BIBentrySTDinterwordspacing}{\spaceskip=0pt\relax}
\providecommand{\BIBentryALTinterwordstretchfactor}{4}
\providecommand{\BIBentryALTinterwordspacing}{\spaceskip=\fontdimen2\font plus
\BIBentryALTinterwordstretchfactor\fontdimen3\font minus
  \fontdimen4\font\relax}
\providecommand{\BIBforeignlanguage}[2]{{%
\expandafter\ifx\csname l@#1\endcsname\relax
\typeout{** WARNING: IEEEtran.bst: No hyphenation pattern has been}%
\typeout{** loaded for the language `#1'. Using the pattern for}%
\typeout{** the default language instead.}%
\else
\language=\csname l@#1\endcsname
\fi
#2}}
\providecommand{\BIBdecl}{\relax}
\BIBdecl

\bibitem{reactNet-conext21}
H.~Mostafaei, S.~M. Miri, and S.~Schmid, ``Reactnet: Self-adjusting
  architecture for networked systems,'' in \emph{Proceedings of the 17th
  International Conference on Emerging Networking EXperiments and
  Technologies}, ser. CoNEXT '21, 2021, p. 473–474.

\bibitem{ray-osdi18}
P.~Moritz, R.~Nishihara, S.~Wang, A.~Tumanov, R.~Liaw, E.~Liang, M.~Elibol,
  Z.~Yang, W.~Paul, M.~I. Jordan, and I.~Stoica, ``Ray: A distributed framework
  for emerging {AI} applications,'' in \emph{13th USENIX Symposium on Operating
  Systems Design and Implementation (OSDI 18)}, Oct. 2018, pp. 561--577.

\bibitem{self-drivingNetwrk-Feamster17}
\BIBentryALTinterwordspacing
N.~Feamster and J.~Rexford, ``Why (and how) networks should run themselves,''
  \emph{CoRR}, 2017. [Online]. Available: \url{http://arxiv.org/abs/1710.11583}
\BIBentrySTDinterwordspacing

\bibitem{programmability-CCR20}
N.~Foster, N.~McKeown, J.~Rexford, G.~Parulkar, L.~Peterson, and O.~Sunay,
  ``Using deep programmability to put network owners in control,''
  \emph{SIGCOMM Comput. Commun. Rev.}, vol.~50, no.~4, p. 82–88, oct 2020.

\bibitem{kalmbach2018empowering}
P.~Kalmbach, J.~Zerwas, P.~Babarczi, A.~Blenk, W.~Kellerer, and S.~Schmid,
  ``Empowering self-driving networks,'' in \emph{Proceedings of the afternoon
  workshop on self-driving networks}, 2018, pp. 8--14.

\bibitem{kellerer2019adaptable}
W.~Kellerer, P.~Kalmbach, A.~Blenk, A.~Basta, M.~Reisslein, and S.~Schmid,
  ``Adaptable and data-driven softwarized networks: Review, opportunities, and
  challenges,'' \emph{Proceedings of the IEEE}, vol. 107, no.~4, pp. 711--731,
  2019.

\bibitem{laghrissi2018survey}
A.~Laghrissi and T.~Taleb, ``A survey on the placement of virtual resources and
  virtual network functions,'' \emph{IEEE Communications Surveys \& Tutorials},
  vol.~21, no.~2, pp. 1409--1434, 2018.

\bibitem{hantouti2018traffic}
H.~Hantouti, N.~Benamar, T.~Taleb, and A.~Laghrissi, ``Traffic steering for
  service function chaining,'' \emph{IEEE Communications Surveys \& Tutorials},
  vol.~21, no.~1, pp. 487--507, 2018.

\bibitem{p4-ccr14}
P.~Bosshart, D.~Daly, G.~Gibb, M.~Izzard, N.~McKeown, J.~Rexford,
  C.~Schlesinger, D.~Talayco, A.~Vahdat, G.~Varghese, and D.~Walker, ``P4:
  Programming protocol-independent packet processors,'' \emph{SIGCOMM Comput.
  Commun. Rev.}, vol.~44, no.~3, p. 87–95, 2014.

\bibitem{sivanathan2018classifying}
A.~Sivanathan, H.~H. Gharakheili, F.~Loi, A.~Radford, C.~Wijenayake,
  A.~Vishwanath, and V.~Sivaraman, ``Classifying iot devices in smart
  environments using network traffic characteristics,'' \emph{IEEE Transactions
  on Mobile Computing}, vol.~18, no.~8, pp. 1745--1759, 2018.

\bibitem{shafiq2016network}
M.~Shafiq, X.~Yu, A.~A. Laghari, L.~Yao, N.~K. Karn, and F.~Abdessamia,
  ``Network traffic classification techniques and comparative analysis using
  machine learning algorithms,'' in \emph{2016 2nd IEEE International
  Conference on Computer and Communications (ICCC)}.\hskip 1em plus 0.5em minus
  0.4em\relax IEEE, 2016, pp. 2451--2455.

\bibitem{nguyen2008survey}
T.~T. Nguyen and G.~Armitage, ``A survey of techniques for internet traffic
  classification using machine learning,'' \emph{IEEE communications surveys \&
  tutorials}, vol.~10, no.~4, pp. 56--76, 2008.

\bibitem{pacheco2018towards}
F.~Pacheco, E.~Exposito, M.~Gineste, C.~Baudoin, and J.~Aguilar, ``Towards the
  deployment of machine learning solutions in network traffic classification: A
  systematic survey,'' \emph{IEEE Communications Surveys \& Tutorials},
  vol.~21, no.~2, pp. 1988--2014, 2018.

\bibitem{frank1994artificial}
J.~Frank, ``Artificial intelligence and intrusion detection: Current and future
  directions,'' in \emph{Proceedings of the 17th national computer security
  conference}, vol.~10.\hskip 1em plus 0.5em minus 0.4em\relax Baltimore, MD,
  1994, pp. 1--12.

\bibitem{MLSurvey-IEEESurvey18}
F.~Pacheco, E.~Exposito, M.~Gineste, C.~Baudoin, and J.~Aguilar, ``Towards the
  deployment of machine learning solutions in network traffic classification: A
  systematic survey,'' \emph{IEEE Communications Surveys \& Tutorials},
  vol.~21, no.~2, pp. 1988--2014, 2019.

\bibitem{challengesofLabeling-NetAI21}
\BIBentryALTinterwordspacing
Y.~Lavinia, R.~Durairajan, R.~Rejaie, and W.~Willinger, ``Challenges in using
  ml for networking research: How to label if you must,'' in \emph{Proceedings
  of the Workshop on Network Meets AI \& ML}.\hskip 1em plus 0.5em minus
  0.4em\relax New York, NY, USA: Association for Computing Machinery, 2020, p.
  21–27. [Online]. Available: \url{https://doi.org/10.1145/3405671.3405812}
\BIBentrySTDinterwordspacing

\bibitem{berry2019supervised}
M.~W. Berry, A.~Mohamed, and B.~W. Yap, \emph{Supervised and unsupervised
  learning for data science}.\hskip 1em plus 0.5em minus 0.4em\relax Springer,
  2019.

\bibitem{ML-CCS22}
\BIBentryALTinterwordspacing
A.~S. Jacobs, R.~Beltiukov, W.~Willinger, R.~A. Ferreira, A.~Gupta, and L.~Z.
  Granville, ``Ai/ml for network security: The emperor has no clothes,'' in
  \emph{Proceedings of the 2022 ACM SIGSAC Conference on Computer and
  Communications Security}, ser. CCS '22.\hskip 1em plus 0.5em minus
  0.4em\relax New York, NY, USA: Association for Computing Machinery, 2022, p.
  1537–1551. [Online]. Available:
  \url{https://doi.org/10.1145/3548606.3560609}
\BIBentrySTDinterwordspacing

\bibitem{int-spec}
{The P4.org Applications Working Group}, ``{In-band network telemetry (INT)
  dataplane specification v2.1},''
  \url{https://github.com/p4lang/p4-applications/tree/master/docs}, 2020.

\bibitem{ML-syscon18}
D.~McGaughey, T.~Semeniuk, R.~Smith, and S.~Knight, ``A systematic approach of
  feature selection for encrypted network traffic classification,'' in
  \emph{2018 Annual IEEE International Systems Conference (SysCon)}, 2018, pp.
  1--8.

\bibitem{ml-classification-HotNet19}
Z.~Xiong and N.~Zilberman, ``Do switches dream of machine learning? toward
  in-network classification,'' ser. HotNets '19, 2019, p. 25–33.

\bibitem{Logstash}
``{Logstash: Collect, Parse, Transform Logs},''
  \url{https://www.elastic.co/logstash/}, 2021.

\bibitem{ffmpeg}
``A complete, cross-platform solution to record, convert and stream audio and
  video,'' \url{https://ffmpeg.org/}, accessed: 2022-04-20.

\bibitem{imageQuality-ICPR10}
A.~Horé and D.~Ziou, ``Image quality metrics: Psnr vs. ssim,'' in \emph{2010
  20th International Conference on Pattern Recognition}, 2010, pp. 2366--2369.

\bibitem{hore2010image}
A.~Hore and D.~Ziou, ``Image quality metrics: Psnr vs. ssim,'' in \emph{2010
  20th international conference on pattern recognition}.\hskip 1em plus 0.5em
  minus 0.4em\relax IEEE, 2010, pp. 2366--2369.

\bibitem{VMAF}
``Vmaf - video multi-method assessment fusion,''
  \url{https://github.com/Netflix/vmaf}, accessed: 2022-04-20.

\bibitem{hoo2017roc}
Z.~H. Hoo, J.~Candlish, and D.~Teare, ``What is an roc curve?'' pp. 357--359,
  2017.

\bibitem{demand-aware-CCR19}
C.~Avin and S.~Schmid, ``Toward demand-aware networking: A theory for
  self-adjusting networks,'' \emph{SIGCOMM Comput. Commun. Rev.}, vol.~48,
  no.~5, p. 31–40, 2019.

\bibitem{zerwas2019netboa}
J.~Zerwas, P.~Kalmbach, L.~Henkel, G.~R{\'e}tv{\'a}ri, W.~Kellerer, A.~Blenk,
  and S.~Schmid, ``Netboa: self-driving network benchmarking,'' in
  \emph{Proceedings of the 2019 Workshop on Network Meets AI \& ML}, 2019, pp.
  8--14.

\bibitem{kim2015band}
C.~Kim, A.~Sivaraman, N.~Katta, A.~Bas, A.~Dixit, and L.~J. Wobker, ``In-band
  network telemetry via programmable dataplanes,'' in \emph{ACM SIGCOMM},
  vol.~15, 2015.

\bibitem{safa-toit23}
\BIBentryALTinterwordspacing
H.~Mostafaei and S.~Afridi, ``Sdn-enabled resource provisioning framework for
  geo-distributed streaming analytics,'' \emph{ACM Trans. Internet Technol.},
  vol.~23, no.~1, feb 2023. [Online]. Available:
  \url{https://doi.org/10.1145/3571158}
\BIBentrySTDinterwordspacing

\bibitem{avin2019renets}
C.~Avin and S.~Schmid, ``Renets: Toward statically optimal self-adjusting
  networks,'' \emph{arXiv preprint arXiv:1904.03263}, 2019.

\bibitem{sleator1985self}
D.~D. Sleator and R.~E. Tarjan, ``Self-adjusting binary search trees,''
  \emph{Journal of the ACM (JACM)}, vol.~32, no.~3, pp. 652--686, 1985.

\bibitem{self-driving-Access21}
K.~Dzeparoska, N.~Beigi-Mohammadi, A.~Tizghadam, and A.~Leon-Garcia, ``Towards
  a self-driving management system for the automated realization of intents,''
  \emph{IEEE Access}, vol.~9, pp. 159\,882--159\,907, 2021.

\bibitem{serviceCoordination-CNSM20}
S.~Schneider, A.~Manzoor, H.~Qarawlus, R.~Schellenberg, H.~Karl, R.~Khalili,
  and A.~Hecker, ``Self-driving network and service coordination using deep
  reinforcement learning,'' in \emph{2020 16th International Conference on
  Network and Service Management (CNSM)}, 2020, pp. 1--9.

\bibitem{verification-NetAI19}
A.~Shukla, K.~N. Hudemann, A.~Hecker, and S.~Schmid, ``Runtime verification of
  p4 switches with reinforcement learning,'' in \emph{Proceedings of the 2019
  Workshop on Network Meets AI \& ML}, ser. NetAI'19, 2019, p. 1–7.

\bibitem{switchML-nsdi21}
A.~Sapio, M.~Canini, C.-Y. Ho, J.~Nelson, P.~Kalnis, C.~Kim, A.~Krishnamurthy,
  M.~Moshref, D.~Ports, and P.~Richtarik, ``Scaling distributed machine
  learning with in-network aggregation,'' in \emph{{NSDI} 21}, 2021, pp.
  785--808.

\end{thebibliography}

\end{document}